\documentstyle[12pt,aasms4]{article}
\slugcomment{Accepted for publication in the November 1997 issue of the Astronomical Journal}

\lefthead{G\'{o}mez et al.}
\righthead{The Dynamics, X-ray Emission, and Radio Galaxies in Abell 578 and Abell 1569}

\begin{document}

\title{The Cluster Dynamics, X-ray Emission, and Radio Galaxies in Abell 578 and Abell 1569}

\author{P. L. G\'{O}MEZ , M. J. LEDLOW, J. O. BURNS, J. PINKNEY\altaffilmark{1},}
\affil{Department of Astronomy,\\ New Mexico State University,\\
    Box 30001/Dept. 4500, \\ Las Cruces, NM 88003-0001,\\
     Email: pgomez@nmsu.edu, mledlow@nmsu.edu, jburns@nmsu.edu, jpinkney@nmsu.edu}

\author{and J. M. HILL }
\affil{Steward Observatory,\\ University of Arizona,\\ Tucson, AZ 85721-0065,\\
     Email: jhill@as.arizona.edu}

\altaffiltext{1}{Current address:  Department of Physics, University of Nevada Las Vegas, 4505 Maryland Parkway, Las Vegas, NV 89154-4002}

% Notice that each of these authors has alternate affiliations, which
% are identified by the \altaffilmark after each name.  The actual alternate
% affiliation information is typeset in footnotes at the bottom of the
% first page, and the text itself is specified in \altaffiltext commands.
% There is a separate \altaffiltext for each alternate affiliation
% indicated above.

% The abstract environment prints out the receipt and acceptance dates
% if they are relevant for the journal style.  For the aasms style, they
% will print out as horizontal rules for the editorial staff to type
% on, so long as the author does not include \received and \accepted
% commands.  This should not be done, since \received and \accepted dates
% are not known to the author.

\begin{abstract}

	We present the results of our multiwavelength study of two nearby Abell clusters that contain extended tailed radio sources. From our analysis of archival PSPC X-ray data, VLA radio observations, and new velocity measurements, we find strong evidence that these clusters are not relaxed. Significant X-ray substructure is revealed by examining deviations from circular models of the overall X-ray surface brightness maps. We report 130 new redshifts from the fields of A578 and A1569 and find that 41 galaxies belong to A578 and 54 belong to A1569. Moreover, we detect the presence of substructure in the velocity and spatial distribution of galaxies. We identify two interacting subclusters in A578 and two gravitationally unbound subclusters in A1569. Furthermore, we find that the radio sources in A578 are at rest with respect to their subclusters and that the Wide-Angled Tailed (WAT) source in A1569 has a small peculiar motion with respect to its subcluster (220 km/s). The standard model of ram pressure induced jet curvature requires relative velocities of the radio sources with respect to the ICM of $\gtrsim$ 1000 km/s. Therefore, the apparent peculiar velocities of the radio galaxies can not explain the bending of the jets/tails for the Narrow-Angled Tailed (NAT) radio source in A578 and for the WAT in A1569. We suggest that a cluster-subcluster merger model provides a reasonable explanation for the X-ray and optical substructure detected in these systems. Numerical simulations of cluster-subcluster mergers predict a bulk flow of ICM gas that has sufficient dynamic pressure for bending and shaping the extended radio sources. Thus, a cluster-subcluster merger can provide the environment responsible for the observed cluster X-ray and optical morphologies and for the bending seen in the radio sources.

\end{abstract}

% The different journals have different requirements for keywords.  The
% keywords.apj file, found on aas.org in the pubs/aastex-misc directory, 
% contains a list of keywords used with the ApJ and Letters.  These are 
% usually assigned by the editor, but authors may include them in their 
% manuscripts if they wish. 

%\keywords{globular clusters,peanut clusters,bosons,bozos}

% That's it for the front matter.  On to the main body of the paper.
% We'll only put in tutorial remarks at the beginning of each section
% so you can see entire sections together.

% In the first two sections, you should notice the use of the LaTeX \cite
% command to identify citations.  The citations are tied to the
% reference list via symbolic KEYs.  We have chosen the first three
% characters of the first author's name plus the last two numeral of the
% year of publication.  The corresponding reference has a \bibitem
% command in the reference list below.
%
% Please see the AASTeX manual for a more complete discussion on how to make
% \cite-\bibitem work for you.   

\section{INTRODUCTION}

	The study of substructure within galaxy clusters provides important constraints for cosmological models
of the evolution of large-scale structures in the Universe. For instance, the amount of substructure detected in a cluster at a given epoch may be used to estimate the mean density of the Universe (e.g., \cite{ric92}). Substructure is believed to be an indicator of an unrelaxed state in which the cluster is in the process of accreting mass via mergers or along large-scale filamentary structure.  Observationally, substructure is detected in the spatial distribution of the cluster gas (i.e., in the X-ray surface brightness) or in the spatial and kinematical distribution of the galaxies. It has also been proposed that the presence of extended radio sources (i.e., Narrow-Angle tailed (NATs) and Wide-Angle tailed (WATs)) is another possible signature of a non-relaxed cluster (\cite{bu94a}, \cite{bur96}) because close interaction between the radio emitting plasma and the turbulent ICM produced by a merger may be responsible for shaping the extended radio sources. In this paper, we test these hypotheses on two nearby Abell clusters that contain extended radio sources.

	Several methods have been developed for identifying and quantifying the amount of X-ray substructure present in galaxy clusters (i.e., \cite{dav93}, \cite{moh93}, \cite{whi94}). In one of the pioneering studies of X-ray substructure, Jones \& Forman (1984) found that up to 60\% of galaxy clusters show evidence of substructure. Numerical simulations of cluster growth and evolution suggest that X-ray substructure originates from hot gas which is responding to a time-evolving gravitational potential (\cite{sch93}, \cite{rot96}, \cite{evr90}, among others). The gravitational potential is continuously changing because dark matter and gas are being accreted through filaments and/or via mergers. Shocks, clumps, and elongations are typical observational characteristics of a cluster merger. Another effect of a merger is to dramatically alter the gas temperature distribution as has been shown in recent ASCA and ROSAT observations (e.g., \cite{rot95}, \cite{mar97}). These results confirm the presence of shocks and temperature asymmetries as characteristics of non-relaxed clusters.

	Optically, a cluster that is still accreting matter and is not virialized may show strong signatures of substructure in the line-of-sight (LOS) velocity distribution and in the spatial (2D) distribution of its galaxies. Numerous statistical tests have been developed for detecting and quantifying optical substructure (see Pinkney et al. 1996 for an overview). 

	Since the evolution of a cluster has significant effects on the intracluster medium (ICM) and on its galaxies, it is tempting to ask if these environmental changes affect the formation and development of radio galaxies. There is some evidence that low power extended radio sources (FR Is) are likely to be found in galaxy clusters that show evidence of substructure in the X-ray and in the optical (\cite{bu94a}). For instance, Pinkney et al. (1997) finds several examples of significant 
spatial-velocity correlations within WAT clusters associated with optically dominant galaxies at the optical/X-ray cluster centers. G\'{o}mez et al. (1997a) studied another sample of WAT clusters and found that  90\% of their sample had significant X-ray substructure and X-ray extensions between the radio tails.  They hypothesized that the X-ray substructure is evidence for recent mergers in these systems, which may explain the characteristic bending of the WAT radio jets. Finally, they proposed that ram pressure caused by the bulk motion flow of gas within a cluster-cluster merger is responsible for re-shaping the radio emitting plasma. 

	Bliton et al. (1997) have recently found a possible correlation between the presence of NATs and X-ray substructure in a sample of optically similar Abell clusters with ROSAT PSPC data. They compared their results with a control sample of radio-quiet clusters and found that NATs are much more likely to be found in clusters with significant substructure. They proposed a new model in which NATs are formed as a result of an increase in nearby ram pressure in cluster-cluster mergers.

	Finally, detailed studies of particular clusters show a very close relationship between the dynamical properties of galaxies, the local ICM, and the properties of extended radio sources. For instance, R\"{o}ttgering et al. (1995) suggested that the radio sources in A2256 (i.e., a radio halo and 5 NATs) were produced by the interaction of a disrupted ICM and the radio emitting plasma. They postulated that the ICM was stirred as a result of an on-going cluster-subcluster merger which was supported by strong optical and X-ray evidence. Moreover, Burns et al. (1995) explained the presence of an extended radio halo and NATs in A2255 as produced by a cluster merger. In their analysis, they compared the observed X-ray and kinematical properties of the cluster with hydro/N-body simulations of a cluster-subcluster merger.
 
	All of these results suggest that there is a correlation between the presence of extended radio sources and the dynamical state of galaxy clusters. We have begun to explore this intriguing possibility by performing a multiwavelength study of a sample of nearby Abell clusters. We are in the process of collecting redshifts and comparing our findings with X-ray images and numerical simulations of cluster evolution. In this paper, we report our results for two of the most interesting members of the sample: Abell 578 and Abell 1569. 

	A578 has 2 prominent extended radio sources: a NAT and a double-lobed (Fat-Double) source (see Figure 1a). It is a relatively poor Abell cluster (richness class RC = 0) and has only 4 published velocities (Struble \& Rood 1992 and a search through the NED archives). A1569 is also a RC=0 Abell cluster that contains a large WAT galaxy and an additional twin-jet source (see Figure 1b). Pinkney (1995) has performed a kinematical study of 22 galaxies with measured redshifts clustered around the WAT source in A1569 and found no significant evidence for non-normality. However, Burns et al. (1994a) and G\'{o}mez et al. (1997a) found significant evidence for substructure from the X-ray surface brightness image. We present new optical and X-ray observations of these interesting clusters, and attempt to understand the relationship between the radio sources and the cluster environs.

	The paper is organized as follows. Section 2 describes the multiwavelength observations. In section 3, we discuss our 
optical and X-ray substructure analysis. In section 4, we analyze the dynamics of both clusters and compare our results with a cluster merger model. 
Finally, we summarize our conclusions in section 5. We use H$_o$=75 km/s/Mpc and q$_o$=0.5 throughout.

\section{OBSERVATIONS}

\subsection {Velocity Measurements}

	Since we are interested in evaluating the dynamical state of these clusters, we need accurate positions
and velocities for the cluster galaxies. Candidates galaxies were identified by eye from the
POSS I plates and their coordinates measured from the
CD-ROM Digital POSS I plates using the guide star plate solutions. The estimated relative accuracy of the digitized POSS I plates is $\sim$ 0\arcsec.5 . We found that
the positions of the radio cores, as measured from the VLA radio maps, agree within 1-2 arcseconds.

	We identified a total of 178 candidate galaxies and 24 very bright stellar objects within $\sim$  one Abell radius
($\sim$ 2 Mpc or 20\arcmin) in the direction of A578 and 70 candidate galaxies and 50 bright stellar objects  within a similar region ($\sim$ 
23\arcmin.5) in the direction of A1569. We observed 123 objects in the direction of A578 and 66 candidate galaxies for A1569. In an effort to estimate the completeness and fairness of our sampling, we have compared the magnitude distribution of the observed objects with the overall distribution of all the candidate galaxies in the respective fields. We obtained rough magnitudes (with an error $\sim$ 0.5) for these objects by using the CD-ROM Digital POSS I plates. We calibrated the candidate galaxies photometry by using the CCD Cousin R-band photometry of the radio galaxies in the clusters (Ledlow \& Owen, 1995b). We find that the A1569 sampling is 90\% complete down to m$_r$=17.5. Our observations of A578 are 80\% complete down to m$_r$=16.5 and  75\% complete down to m$_r$=17.0.

	All the spectra were obtained in 4 observing runs with the MX multifiber spectrometer mounted on the 2.3m Bok Telescope at the University of Arizona Steward Observatory. Table 1 shows a log of the observations. The MX multifiber spectrometer has 32, 2\arcsec ~fibers covering a field of view of 45\arcmin . There are 28 additional fibers used for obtaining sky spectra. The spectrometer has a 400 groove mm$^{-1}$ grating which provides a dispersion of 2.8 \AA  ~ per pixel$^{-1}$ and a spectral coverage from 3700 to 6950 \AA ~ on an UV flooded Loral 1200x800 thinned CCD. A more detailed
description of the instrument can be found in Hill \& Lesser (1988).

	We reduced the observations by following a very similar procedure to the one described by Hill \& Oegerle (1993). We constructed a sky spectrum by combining $\sim$ 28 sky spectra and then we subtracted this spectrum from each galaxy spectra. The scaling of the sky with respect to the galaxy spectra is based on the residual flux at the [OI] sky emission-line ($\lambda$5577). Radial velocities were determined by using the IRAF task `FXCOR' to cross-correlate the observed spectra with 20 templates. We used a similar criterion as the one described by Pinkney et al. (1994) to identify the best heliocentric velocity for the galaxies.

	Tables 2A and 2B contain the measured velocities for each cluster. The first column lists the galaxy ID number. Columns 2 and 3 list the epoch 2000 coordinates. Columns 4 and 5 list the heliocentric velocities and errors. The errors were computed from a fit to the Tonry \& Davis (1979) `R' parameter (TDR) vs. velocity errors obtained from redundant MX measurements of galaxy velocities (\cite{pin97}). Column 6 shows the TDR values obtained from the cross-correlations. We have excluded spectra of very low S/N objects that had a low TDR number ($<$ 3.5) or those that were clearly stars. We chose this TDR $>$ 3.5 value cutoff because Pinkney et al. (1994) found that measurements with these TDR values would be generally good estimates of the redshift ($\sim$ 74\% of correct cross-correlation peaks, Pinkney, private communication). Finally, the results of the cross-correlations were confirmed from visual inspection of the spectra.
		
\subsection {X-ray Observations}

    The X-ray data for these clusters were obtained from the ROSAT PSPC pointed observations archive. A preliminary X-ray analysis of A1569 was reported by G\'{o}mez et al. 1997a. We have performed a similar analysis for the A578 X-ray data. We used the Snowden et al. (1994) method because we are interested in the extended cluster X-ray emission. This method identifies four main contaminants within the X-ray data:  Particle-induced Background (PB), After-pulse Background (AP), Scattered Solar X-ray Background (SB), and  Long-term Enhancement Background (LTE). Next, it models each contaminant and corrects for them in each one of the Snowden ROSAT bands (R4, R5, R6, and R7) which roughly corresponds to the spectral range from 0.5 keV to 2.05 keV. Moreover, this method accurately creates exposure maps for every band. The final images consist of the summed maps over the range R4-R7. 

\subsubsection {Temperature Fits}
	The spectral analysis of the X-ray data was performed by using the `XSPEC', `FTOOLS', and `XSELECT' software in a similar fashion as in G\'{o}mez et al. (1997a). First, we extracted two spectra from the data files with the same time filters as the ones used for the spatial analysis. One spectrum corresponds to the source region (with point sources removed) while the other spectrum corresponds to the background region. All the source regions had a radius of 500 kpc. Table 3A shows the approximate center and radius of the
source regions (500 kpc), and the inner radius and width of the background annuli.

 We rebinned the counts in the spectra such that every channel had at least 30 counts. Next, we fit the spectral profiles to Raymond-Smith (\cite{ray77}) emission 
models (bremsstrahlung plus lines) in the energy range from 0.5 keV to $\sim$ 
2.05 keV. We assumed negligible 
internal absorption and a fixed cosmic abundance of 0.3 solar which is consistent with the 
values found by Edge \& Stewart (1991a) for a sample of bright clusters observed 
by EXOSAT (their mean cosmic abundance is 0.32 with an rms of 0.18). We checked the quality of our 2 parameter fits (T and $N_H$) by comparing the
resultant hydrogen column density with values obtained from a map of galactic HI
(provided by F. J. Lockman, see \cite{sta92}). We found the fitted values 
to be in close agreement with those in the HI map. We then proceeded to 
fix N$_H$ (Table 3B, third column)  to the Stark et al. (1992) values and fit only
the temperatures with `XSPEC.'

	In Table 3B, we report the results of the temperature fits.
The fourth column lists the 
fitted temperature and the associated errors at the 68\% confidence level for the
$\chi^2$ statistic. The fifth column lists
the reduced $\chi^2$ from the fits and the number of degrees-of-freedom. The sixth column shows the temperatures from the 
C-statistic fits. These fits are based on the Maximum Likelihood method (Cash 1979); thus, they are considered more robust than the common $\chi^2$ fits in cases in which the statistics are low for weak sources. 

\subsection { X-ray/Radio Overlays}

     In order to compare the radio and X-ray morphologies, we first constructed final X-ray images 
in the range 0.5 keV to 2.05 keV. Then, we processed these images with an adaptive kernel smoothing algorithm (\cite{hua97}) which readjusts the size of the smoothing length across the image to simulate a constant signal-to-noise ratio ($\sim$ 4) across the map. Next, we re-gridded and repositioned the X-ray
images to match the VLA maps using the AIPS task HGEOM. These grids contained the same
pixel size, field of view, and image center as the VLA maps. Grey-scale images of the radio
emission were then overlaid onto the contours of the X-ray surface brightness. These X-ray/radio overlays are displayed in Figures 1a and 1b.

\section {ANALYSES OF THE DATA}

\subsection {Optical Analysis}

	The analysis of the optical data consisted of the following steps. First, we determined the galaxy memberships in each cluster. Next, we performed 1D, 2D, and 3D statistical tests in order to look for evidence of substructure. If we detected substructure within the cluster, we would 
attempt an  {\it objective} partition of the galaxies and perform statistical tests on the resulting galaxy subgroups.

	In order to determine the cluster memberships, we first computed the biweight estimators for the scale (S$_{BI}$) and location (C$_{BI}$) (\cite{dan80}, \cite{bee91}) of the velocity data. The main motivation for using these biweight estimators, which are analogous to the classical mean and dispersion, is that they are statistically robust. We computed the estimators errors by using the jacknife technique. Next, we applied the usual 3$\sigma$-clipping criteria (\cite{yah77}) for establishing the galaxy memberships in each cluster.

	Once we were reasonably confident that we had eliminated most of the obvious interloper galaxies, we tested the data for substructure. For this task, we used Pinkney et al.'s (1996) {\it BATCH} program which performs a battery of 1D, 2D, and 3D tests on the spatial and kinematical data. The 1D section of the program includes the A, U, W, B1, B2, B1-B2, I, DIP, KS, V, W$^2$, U$^2$, A$^2$, TI,  and AI-tests. Most of the 1-D tests compare the velocity distribution with a Gaussian distribution. Thus, the significance of the test reveals the probability that the two distributions are consistent; i.e. a low value for the probability means that the distribution is significantly non-Gaussian. Other tests analyze the shape of the distribution (i.e., TI test). The software also provides an average of the probabilities for non-normality, skewness, and kurtosis. 

	The 2D tests are the Angular Separation Test (AST), the Fourier Elongation Test (FEL), the $\beta$ test, and the 2D Lee statistic. The 2D Lee statistic uses reshuffling of the galaxy positions to estimate the significance of any substructure. Therefore, a low probability indicates significant substructure. 

	Finally, the program uses the following 3D tests: Lee 3D, $\Delta$ (Dressler-Shectman or DS), $\epsilon$, the $\alpha$ tests (West-Bothum or WB), and the $\alpha$ variation test (WBv). The significance of these tests is determined by using a Monte-Carlo method. Thus, they test how many times the observed measurement is greater than the simulations based on the reshuffling of the observed data. This implies that low probability numbers identify significant substructure.

	Whenever we detected statistically significant substructure, we used the KMM (\cite{bir92}, \cite{bir93}, \cite{dav95}) algorithm to objectively partition the data in sub-groups. Briefly, this algorithm models the velocity distribution and the projected spatial distribution of the galaxies as composed by a user-defined number of sub-groups. This is the so called {\it mixture modelling} technique. Each one of the sub-groups is in turn modelled as a Gaussian. The program finds the best fit by assigning different galaxies to different groups until the likelihood statistic is maximized. This program needs a first guess for the centroid (3D) position of the subgroups, their relative sizes (3D), and their estimated errors. As an additional check, the method also provides an estimate of the formal probability of rejecting the hypothesis that one Gaussian would fit the data.

\subsubsection {A578 Galaxy Velocities}

	In Figure 2a we show a histogram of the velocity distribution for the galaxies measured in this field. The main cluster of 41 observed galaxies (which includes the two radio galaxies of interest) is located at $\sim$ 27,000 km s$^{-1}$ while a background group of 17 galaxies is located at $\sim$ 40,000 km s$^{-1}$. We have identified the NAT radio galaxy 0720+670 with our galaxy \#5 and 0719+670 with galaxy \#1 in Table 2A (\cite{owe97}). The values for the biweight estimators and other data for A578 are listed in Table 4. 

	Figure 2b shows a velocity-coded 2D plot of the spatial distribution of the galaxies in this field. The background galaxies appear to be uniformly distributed spatially whereas the galaxies clustered at 27,000 km s$^{-1}$ are more centrally concentrated. 

	Figure 2c shows a histogram of the velocity distribution of the accepted galaxies for A578 (these are all the observed galaxies clustered at 27,000 km s$^{-1}$ that survived the 3$\sigma$ clipping). This distribution does not appear to be Gaussian. In order to quantify the non-normal character of this distribution and to detect other signs of substructure, we ran the statistical tests described by Pinkney et al. (1996). Table 5 shows the results of the substructure analysis. The 1D tests indicate possibly significant kurtosis (average = 9\%) and a significant asymmetry (AI=5\%). Although there is no evidence for substructure in the 2-D spatial distribution of the galaxies, we find evidence for significant 3D substructure (2\%) from the $\alpha$ test. Figure 2d shows the existence of an apparent spatial segregation between the low velocity galaxies and the high velocity galaxies. One group of galaxies is concentrated in the South with velocities $\leq$ 27,000 km/s while the other galaxies with velocities $\geq$ 27,000 km/s are predominantly located in the North. It is surprising, however, that the other 3D tests do not find any significant evidence of substructure. However, Pinkney et al. (1996) pointed out that these 3-D tests have widely varying sensitivity to different substructure dependent on the viewing angle relative to the merger axis and the epoch of the merger.

	Because we found significant 1D substructure and possible 3D substructure, we used the mixture modelling technique in order to identify the subgroups present in the data. Table 6A shows the results of applying the KMM algorithm to A578. The method splits the galaxy population into two groups that we labeled  North (18 members) and South (23 members) due to their centroid positions. Most of the galaxy assignments which are listed in Table 6B had a formal probability $>$ 95\% to belong to their listed subgroups. Only galaxies \#15, \#20, and  \#27 had marginal probabilities $>$ 80\%. The observed distribution is consistent with a single Gaussian distribution at the very small 0.008\% level. This significance might not mean that each galaxy assignment is correct {\it per se}, but it justifies our effort to characterize the components responsible for the substructure. To verify that the results were reliable, we performed the same analysis by changing the initial guesses for the sub-groups centroids and errors as suggested by Davis et al. (1995). The algorithm converged to the same results regardless of the guess for the initial parameters. Figure 2e shows a sub-group coded 2D spatial distribution of the galaxies from the KMM partitioning. It is reassuring to note that the new
assignments show a distribution that is very similar to the naive velocity-based splitting depicted in Figure 2d. 

	Figure 2f shows two contour plots overlaid onto a grey-scale image of the cluster X-ray emission. The contour plots represent the galaxy surface density distribution of the two subgroups identified by the KMM algorithm. Both density distributions were smoothed with an adaptive kernel algorithm described by Beers et al. (1991). We see several interesting correlations between the X-ray and optical data. First, both galaxy subgroups appear to be elongated. The southern subgroup shows a E-W elongation while the northern subgroup exhibits a N-S elongation. Second, these elongations in the galaxy distribution appear to trace the cluster X-ray emitting gas. Finally, the centroid of the southern group is centered on a X-ray peak while one of the secondary peaks of the northern group is located close (1\arcmin.2 or 110 kpc) to a local peak of X-ray emission.
 
	Where are the radio galaxies located? Galaxy \#1 is located at the center of the southern group and the NAT is located close to one of the subpeaks of the southern group (compare Figures 1a and 2f). Kinematically, they appear to be at rest within their subgroups. We estimated the significance of the radio galaxy velocities using the analysis of Teague et al (1990). They introduced the $S_V$ statistic defined as follows:

\begin{equation}
\ S_V = \frac{| C_{BI} - v|}{(\sigma_1^2 + \sigma_2^2)^{\frac{1}{2}}}
\end{equation}
\noindent
where {\it v} is the galaxy velocity and $\sigma_1$ and $\sigma_2$ correspond to the errors in the cluster location and galaxy velocity. A value of $S_V <$ 1 would mean that the galaxy is at rest and $S_V >$ 2.0 corresponds to a marginally significant peculiar motion. When we analyze the kinematics of the radio sources with respect to the overall cluster, we find that galaxy \#1  (v$_{pec}$=622 km/s and $S_V$= 4.0) and galaxy \#5 (v$_{pec}$=935 km/s and $S_V$= 6.0) have significant peculiar velocities. However, after we correct for the substructure present in the systems, we find that these radio galaxies are in fact at rest with respect to their subgroups (i.e., $S_V <$ 1 when we only consider the velocities within each subgroup). This poses an interesting question:  how is the southern tailed source bent if the radio galaxy is at rest?

\subsubsection {A1569 Galaxy Velocities}

	Figure 3a shows a histogram of the velocity distribution for the galaxies measured in A1569. Almost all of the observed galaxies are clustered at $\sim$ 22,000 km s$^{-1}$. There are only 2 galaxies located outside this velocity range ($\sim$ 31,500 km s$^{-1}$). We have identified the two radio sources associated with A1569 with the galaxies clustered at $\sim$ 22,000 km s$^{-1}$. Galaxy \#1 corresponds to 1233+169 and galaxy \#2 corresponds to the WAT radio source 1233+168 (\cite{owe97}). The values for the biweight estimators are listed in Table 4. Note the unusually high scale value ($\sim$ 1618 km/s) for a richness class 0 cluster. 

	Figure 3b shows a histogram of the velocity distribution of galaxies centered at 22,000 km s$^{-1}$. This distribution is clearly non-Gaussian and shows the presence of at least two groups of galaxies. One relatively compact group appears to be clustered at $\sim$ 20,700 km s$^{-1}$ and the other more dispersed group is concentrated at $\sim$ 24,000 km s$^{-1}$. This explains the very large value of the biweight scale for this cluster. However, are these two groups also separated spatially? In Figure 3c we show a velocity-coded 2D spatial distribution of the galaxies which indicates that the two groups are indeed spatially segregated. The low velocity galaxies are concentrated in the S-W area near 1233+168, whereas the high velocity galaxies are concentrated towards the N-E area near 1233+169.
 
	 As expected, the statistical analysis finds strong evidence for substructure (Table 7). There is significant kurtosis ($<$ 0.1\%) and non-normality (3\%). Moreover, the TI and AI reflect the velocity contamination present due to the N-E group. There is also evidence for 2D substructure. The $\beta$ test reveals the asymmetry in the distribution of galaxies and the Lee 2D test points towards the apparent bimodality of the spatial distribution of the galaxies. Furthermore, the Lee 3D and $\epsilon$ tests also show evidence for substructure.

	The KMM algorithm successfully partitions the galaxies into two groups. The basic data for the subgroups and the galaxy assignments are listed in Table 8A and Table 8B. The mixture modelling technique assigns 25 objects to the N-E group and 29 objects to the S-W group (Figure 3d). Galaxy \#51 had the lowest probability of group membership (99\%). Moreover, the method finds that the distribution is consistent with a single Gaussian at $<$ 0.1\% level. Kinematically, the KMM method also separates the galaxies based on their velocities (Figure 3e). The galaxies located at the N-E have the highest velocities whereas the galaxies at the S-W have the lowest velocities. 

 	Figure 3f shows an overlay of the X-ray emission from the cluster (gray scale) onto a contour plot of adaptively smoothed galaxy surface density. As for the case of A578, we find that the galaxy density peaks are coincident with X-ray clumps and that both maps show similar elongations. Interestingly, the two subgroups appear to be connected by gas and by a tenuous trail of galaxies.

	We find that the WAT galaxy has a marginally significant ($S_V$=2.2) peculiar motion (215 $\pm$ 105 km/s) with respect to the dynamical center of its subgroup. Note that this velocity, however, is insufficient to account for the bending of the jets.

\subsection {X-ray Analysis}

     Another probe of the dynamical state of the cluster is the X-ray emission from its intracluster gas.
Any substructure in the gas is probably caused by an evolving gravitational potential or deviations from hydrostatic equilibrium. Moreover, a study of the physical properties of the gas will allow us to determine if the cluster gas and the radio emitting plasma are in pressure equilibrium.

\subsubsection {X-ray Substructure}

	Figures 1a and 1b reveal a very interesting spatial correlation between the position of the 
extended radio sources and the presence of X-ray clumps for our two clusters. In these Figures we
note that all the radio sources appear to coincide spatially with clumps of X-ray emission. This effect was first noted by Burns et al. (1994a) from lower resolution {\it Einstein} images.
In order to quantify the significance of these X-ray clumps, we performed a substructure analysis of the
X-ray emission from these clusters. 

	 Our analysis for A1569 has been previously presented in G\'{o}mez et al. (1997a) and is similar to the method used by Davis \& Mushotzky (1993). In short, our method identifies substructure by comparing the actual X-ray emission from the cluster with a circular model. Next, we estimate the significance of the substructure by using Poisson statistics. We have removed any obvious point sources during our analysis. Therefore, all the structure that we list in Table 9 as significant is not produced by point sources.

\subsubsection {Intracluster Gas Properties}

	In order to characterize the general morphology of the clusters, we fitted $\beta$  models (e.g., \cite{sar86}) to the background subtracted profiles. The profiles were generated by azimuthally averaging
concentric annuli with a bin size = 30\arcsec ~using Poisson statistics for the errors (background + sources counts added in quadrature). We used circular annuli in order to create the profiles and masked all the point sources in the region of interest.
The best fit values for A578 are $\beta$= 0.6 $\pm$ 0.1, a core radius of 120 $\pm$ 8 kpc, with a reduced $\chi^2$= 0.6. For A1569, G\'{o}mez et al (1997a) report values of $\beta$= 0.6, a core radius of 182 $\pm$ 110 kpc, and a reduced $\chi^2$= 0.5. Although the quality of the fits are marginal, these fits are a convenient parameterization of the cluster properties.

     We used the deprojection technique of Arnaud (1988) to estimate the density, temperature, and pressure profiles for the cluster gas. This method assumes a Raymond-Smith (1977) thermal spectrum, that the gas is in quasi-hydrostatic equilibrium, and requires an initial estimate for the temperature in the
outermost boundary of the profile and the scale of the X-ray emission from the $\beta$ model fit ($\beta$ and $r_c$).

	The deprojection technique also needs an estimate for the velocity dispersion. In our case, we replaced the velocity dispersion by the biweight estimator of the scale. We used 380 km/s for A578 (southern clump of galaxies) and 433 km/s for A1569 corresponding to the southern X-ray clump. The densities determined from the deprojections, along with the measured temperatures,
were used to estimate the ICM pressures listed in Table 10.

	Finally, we compared the thermal pressures from the deprojection to the minimum pressure
in the radio tails (e. g., \cite{ode85}). We used the radio images from the VLA 20-cm radio survey of Abell clusters (\cite{owe97}) for this analysis. The minimum internal pressures were measured at the middle of the diffuse portions of the radio tails. 
For our calculations, we assumed a filling factor of 1, a power-law spectrum with $\alpha=1$ ($F_\nu \propto \nu^{-\alpha}$) with lower and upper cutoffs of 10 MHz and 100 GHz, a transverse magnetic field ($sin\theta = 1$), cylindrical symmetry, and an approximate equipartition of energy between the relativistic electrons and protons.

	Table 10 lists the results for each cluster. The tails of the two radio sources in A578 appear to be in rough pressure equilibrium (within a factor of 2). However, the thermal and radio plasma pressures for A1569 are quite different. The tails of the southern WAT appear to be underpressured by more than a factor of 10 with respect to the surrounding ICM. A similar trend has been found in other cluster radio sources such as A2255 (\cite{bur95}), Hydra A (\cite{tay90}), and a sample of cluster sources (\cite{fer92}). The presence of thermal gas accreted from the ICM in the tails (i.e., entrainment) has been suggested as an explanation for this effect because the entrained intracluster gas will contribute its own thermal pressure inside the radio tails which is not taken into account by the minimum pressure computation described above.

\subsubsection {Jet Bending}

	In this section, we discuss a feasible mechanism for the bending of the jets based on the O'Dea (1985) and Doe et al. (1995) analysis of cluster NAT radio sources. In their studies, they modeled the NATs as `cold' jets (Mach numbers $\gg$ 1) that are bent due to ram pressure from interaction with the ICM. Thus, they assume that the jet can be modeled as a laminar relativistic fluid that obeys the relativistic and time independent Euler's equation

\begin{equation}
\ \frac{{\rho_j}{{v_{j}}^2}{\gamma^2}}{R}\approx \frac{{\rho_{ICM}}{{v_{rel}}^2}}{h} ,
\end{equation}

\noindent
where $v_j$ is the jet velocity, $\rho_j$ is the jet density, $\beta_j = v_j/c$, $\gamma=(1-\beta_j^2)^{-1/2}$,  R is the radius of curvature of the jet, $v_{rel}$ is the relative velocity of the radio source with respect to the ICM,$ h $ is the scale length over which the ram pressure is effective, and $\rho_{ICM}$ is the gas density.

Next, they assumed that the initial jet kinetic energy is converted to the observed radio luminosity with an efficiency of $\epsilon$. This yields

\begin{equation}
\ v_{rel} \approx \sqrt {\frac{ L_{rad}\beta_{j}\gamma h}{(\gamma-1) \rho_{ICM} \epsilon \pi {r_j}^2 c R} (\frac{r_{ji}}{r_{if}})^{2/3}} ,
\end{equation}

\noindent
where $r_j$, $r_{ji}$ and $r_{if}$ are the average, initial, and final jet radius and $L_{rad}$ is the radio luminosity. As mentioned by Doe et al. (1995), this expression is very useful because it involves quantities that are easily determined from the multiwavelength observations of the clusters ($r_j$, $\rho_{ICM}$, L$_{rad}$). However, we still have to make some assumptions for $\epsilon$ and $\beta_{j}$.

	As a first step in our analysis, we calculated the radio luminosities for the jets in A578's NAT and the A1569's WAT. These luminosities were computed from the flux densities by assuming a spectral index of 0.5 for the jets and a bandwidth from 10 MHz to 15 GHz as in O'Dea and Owen (1987). We obtained $L_{A578}$=0.83 x $10^{41}$ erg/s and $L_{A1569}$=4.8 x $10^{41}$ erg/s.

	Figure 4 shows plots of $v_{rel}$ and jet number density as a function of $\beta_j$ for the southern A578 NAT radio source and for the southern A1569 WAT radio source. This Figure illustrates that the jet is more difficult to bend at high densities and/or if the jet is slow. On the other hand, a fast and/or tenuous jet is easier to bend. 

 In order to further constrain the different parameters governing the jet bending, we have examined numerical simulations of the disruption of jets (\cite{zha92}); these simulations show that a jet would be disrupted at a characteristic jet disruption length which is a function of the jet Mach number. By examining the radio data from these sources, we have determined that both jets are disrupted very close to their host galaxy. Thus, it is safe to assume that the jets propagate without disruption from some 20-30$r_j$ (these are their jet disruption lengths). This will imply Mach numbers between 4 and 5 (see Figure 7 in Zhao et al. 1992). Moreover, if we assume that the typical sound speed of the ICM surrounding the jets is $\sim$ 500 km/s, we find $\beta_j$ values in the range  $ \sim 0.007-0.008$. These $\beta_j$ values constrain the possible relative velocities (Figure 4) between 400 - 2500 km/s. {\it This is the range of relative velocities of radio sources with respect to the ICM that is responsible for the observed radio morphology}. But, how could these high relative velocities arise if the radio galaxies are at rest with respect to the other galaxies in their subgroups (see section 3)?. We will present a possible solution to this puzzling problem in the next section.

\section { DISCUSSION}

     Our analysis of the X-ray and optical data confirms our suspicions that these clusters are not relaxed systems. It is important to note that we initially chose these clusters only because they contain extended radio sources. In what follows, we describe our general impressions for each cluster and discuss models for the jet bending.

\subsection {A578}
	
	From our multiwavelength observations of the A578, we find significant substructure in the gas density, as revealed by the X-ray surface brightness maps, and in the distribution of the cluster galaxies. When we compare the results of our X-ray and optical analysis, we discovered very interesting properties. First, the velocity dispersion of the whole system (793 km/s, Table 4) does not agree with the overall X-ray temperature of the gas. We estimate that a velocity dispersion of 500 km/s would be in better agreement with the X-ray temperature of this cluster according to the T-$\sigma$ relationship for clusters of galaxies (\cite{ed91b}). Interestingly, the optical subgroups identified by the KMM algorithm have velocity dispersions ($\sim$ 400-500 km/s) that are in much better agreement with the cluster X-ray temperature. Second, these galaxy subgroups appear to coincide spatially with X-ray clumps (Figure 2f).

	We believe that these correlations and the other cluster properties that we have described in previous sections indicate that this cluster has undergone a recent cluster-subcluster merger. This merger will not only explain several of the X-ray and optical properties in this cluster, but it will also offer a solution to the puzzling problem of the NAT bending.

	In order to explore this merger model further, we have performed numerical simulations of cluster-subcluster mergers and compared them with our observations. The simulations were generated with a code that is a hybrid between the Hernquist's TREECODE and the NCSA's ZEUS-3D hydrodynamics code. Thus, this program is very similar to the one used by Roettiger et al. (1996). The main differences are the use of a new Poisson solver for the creation of the gravitational potential, the inclusion of radiative cooling in the treatment of the ICM energy budget, the use of symmetry along the merger axis which allowed us to obtain better spatial resolution (20 kpc), and the ability to trace the gas in each subcluster with the help of passive scalars (\cite{go97b}).

	These simulations start with two idealized clusters (main cluster and a subcluster) that are modeled as isothermal King spheres separated by $\sim$ 4 Mpc that are allowed to interact under the effects of their own gravity. As the merger progresses, we observe the typical characteristics of merging clusters. For instance, we find that the gas is heated, compressed, and develops shocks. Moreover, the gas becomes clumpy and elongated as it tries to regain hydrostatic equilibrium with an evolving gravitational potential. All of these properties are similar to the ones identified by other numerical simulations of cluster mergers (e.g., \cite{rot96}, \cite{sch93}). 

	In order to compare these simulations with the actual data, we identified the simulation that most closely resembled the observed X-ray and optical morphology of the cluster. A merger between two subclusters with mass ratio of 4 to 1 about 1 Gyr after core crossing is our best matching simulation (compare Figure 1a  and 2f with Figure 5). Figure 5 shows a map of the logarithm of the simulated gas density. We have also overlaid onto this Figure two contour maps which represent the surface density of the N-body particles (dashed contours). One of the contour plots shows the particles belonging to the main cluster (short dashes) while the other contour plot shows the distribution of the secondary cluster particles (long dashes). If we assume that the N-body particles represent the cluster galaxies, we find very interesting morphological similarities. A close comparison between this simulation and the observations can explain several of the cluster properties.

	First, it will explain the apparent spatial, but not kinematical, coincidence of two distinct groups of galaxies. During this merger, especially if viewed close to the line of sight, one would expect that the galaxies of each pre-merger cluster would appear mixed in the plane of the sky. However, they would be separated in velocity space, as is the case for this cluster.

	Second, the X-ray substructure that we detect in the system can be explained by the elongations and clumpiness observed in the simulated gas density. These deviations from a relaxed, symmetric, and circular distribution are evidence of a system that is not in hydrostatic equilibrium. Evidence of this effect has been found in other samples of non-relaxed clusters (e.g., Mohr et al 1995). A similar result has been found in the analysis of synthetic X-ray images from numerical simulations of cluster mergers  (\cite{rot96}).

	Third, this simulation shows that gas belonging to the infalling subcluster is protected, and is not completely stripped by ram-pressure. The protective mechanism is the formation of a bow shock that surrounds the subcluster (\cite{rot96}). This effect might explain the fact that the southern radio source is located within a region of X-ray emitting gas. In this model, the northern radio source is associated with the main cluster and hence to the main clump of X-ray emitting gas. 

	Fourth, the coincidence between the position of the N-body particles with the position of the subcluster galaxies can explain the fact that the radio galaxies appear at rest with respect to their local subcluster.

	Fifth, we can explain the bending of the southern NAT from the interaction between the radio jets with the dynamic ICM. In this model, the radio galaxy is located at rest with respect to its subgroup which is consistent with the optical data. However, the relative velocity of the ICM with respect to the galaxy could be large ($\sim$ 1000 km/s) as a result of a bulk flow of gas caused by the merger.

	Figure 6 shows a histogram of the relative velocity of N-body particles with respect to the gas velocity for the 4 to 1 merger simulation. For this plot, we have chosen the N-body particles located within the left clump of the secondary cluster in Figure 5. Next, if we assume that the N-body particles represent the cluster galaxies, Figure 6 shows the actual relative velocity of the subcluster galaxies surrounding the NAT with respect to the cluster ICM (again, compare Figure 1a with Figure 5). It is interesting to note that these velocities are distributed from $\sim$ -1500 km/s to 2000 km/s. These are the velocities that are required to explain the jet bending (section 3.2.3). Therefore, despite the fact that the radio galaxy is at rest with respect to the other galaxies, the radio galaxy will not be at rest with respect to the local ICM. It is important to point out that we do not consider this simulation as a unique model for A578. We are only using these simulations as guidelines for the type of gas dynamics that are produced during cluster mergers.

\subsection {A1569}

	At first glance, A1569 appears to be a bimodal X-ray cluster. The clumps of X-ray emission are associated with each of the two radio sources. Moreover, we have also identified two distinct groups of galaxies (N and S) also associated with the X-ray clumps and the radio galaxies. 

	Are these two groups of galaxies gravitationally bound?. The velocity dispersion of the whole system ($S_{BI}$= 1618 km/s) appears to be too large for the X-ray luminosity and for the overall optical richness of the cluster. We have applied a 2-body analysis to this system in order to determine if there is a projection angle that would find a bound solution for the system. In this analysis, we use the Newtonian criteria for gravitational binding (\cite{bee82}) and the binding ratio as described by Davis et al. (1995). A binding ratio $\leq$ 0.39 will imply that there is a projection angle that will provide a bound solution for the system. In our case, we obtained a binding ratio of 2.06 which reveals that there is no bound solution for the system. It is important to note that we used mass estimates for the systems derived from the optical data and not from the X-ray observations during our analysis (Virial Mass for the northern group = 3.4 x 10$^{14} M_{\sun}$, southern group = 1.1 x 10$^{14} M_{\sun}$). The system will be bound only if the mass estimates for the subgroups are gross underestimates of the real values (at least one order of magnitude) which seems unlikely. 

	Once we divided the galaxies in these two groups, we concentrated our study on the southern group. Our analysis of the southern group of galaxies reveals important kinematical and X-ray properties. As is the case for A578, we consider that a similar merger model provides the best explanation for those properties. 

	First, this model provides an explanation for the peculiar motion of the WAT. It has been suggested in the past (\cite{oge94}, \cite{pin95}) that large peculiar motions observed for massive galaxies, especially for a WAT host, are the result of velocity contamination produced by a merger.

	Second, our X-ray morphological tests find evidence of substructure. Again, a look at Figure 5 can provide a reasonable explanation for the elongation and clumpiness detected in A1569's southern clump (Figure 1b). This substructure suggests that the gas is not in hydrostatic equilibrium.

	Third, a merger model will also explain the similar morphologies between the spatial galaxy distribution and the X-ray emission (Figure 3f). Both distributions exhibit a E-W elongation, which is a possible signature of a merger (see Figure 5). 

	Fourth, we find that the WAT peculiar velocity is 215 $\pm$ 105 km/s. This velocity is not large enough to bend the WAT since $\sim$ 1000 km/s is required according to our models (section 3.2.3). The observed WAT peculiar velocity is probably the result of velocity contamination produced by the galaxies from a merging subcluster. This merger could also explain the WAT bending as produced by the interaction between the dynamic ICM and the jet tails. This mechanism is similar to the one described for A578. Finally, G\'{o}mez et al. (1997) points out that there is a correlation between the local X-ray elongation and the direction of the bending of the jet tail. They explained that this alignment is another signature of the effects produced by a merger that bends and re-shapes the radio emitting plasma.

\section {SUMMARY AND CONCLUSIONS}

	We obtained 74 new redshifts in A578 and find that 41 galaxies are cluster members. We also gathered 56 new velocity measurements in A1569 and confirmed that 54 of those galaxies are cluster members. Our analysis of these optical data, X-ray archival data, and radio data has revealed the following properties:

	First, both clusters show evidence of X-ray substructure. The significant substructure was detected as deviations from circular models of the cluster X-ray emission (clumps and elongations). The X-ray emission from A578 shows two elongations; one is in the N-S direction while the other is in the E-W direction (Figure 1a). A1569 appears to be a bimodal X-ray cluster with an elongated (E-W direction) southern clump (Figure 1b).

	Second, our 1-D, 2-D, and 3-D statistical tests find optical substructure in both clusters. The KMM algorithm successfully identifies two subgroups of galaxies present in these sytems. Thus, both clusters are divided into two subgroups each. Each subgroup contains one extended radio source.

	Third, the positions of the galaxy subgroups coincide with the position of the local X-ray clumps. This seems to suggests that these subgroups are composed of galaxies and gas. A more detailed analysis reveals that the two subgroups in A578 are bound and coincide spatially but not kinematically. On the other hand, our 2-body analysis of A1569 finds that its two subgroups are not gravitationally bound.  

	Fourth, it is interesting that the southern NAT in A578 appears to be at rest with respect to its subgroup galaxies. However, our model of ram induced jet bending requires a relative velocity of the jet with respect to the ICM of $\gtrsim$ 1000 km/s in order to reproduce the observed morphology. So, what causes the NAT bending?

	Fifth, the WAT in A1569 exhibits a small peculiar velocity ($\sim$ 200 km/s) which is not enough to explain the bending of the WAT according to a ram-pressure model.

 	We believe that a cluster-subcluster merger model provides a reasonable explanation for the observed optical and X-ray properties of these clusters and for the bending of their radio sources. During a merger, the cluster gas will be disrupted and shocked as it tries to regain hydrostatic equilibrium. Morphologically, the gas will show clumps and elongations that are similar to the ones seen in our observations of these clusters. This dynamic environment also produces bulk flows of gas ($\gtrsim$ 1000 km/s) that in turn are responsible for the bending of the radio tails. Thus, we believe that a merger is the best mechanism to explain the presence of extended and bent radio sources in these clusters. Therefore, we have confirmed our suspicions that extended radio sources act as beacons of merger activity in these two clusters.

	Finally, we would like to point out that these two clusters were chosen simply because they contained galaxies with extended radio emission and had available ROSAT PSPC data. We are in the process of performing a similar analysis on a larger sample ($\sim$ 15) of galaxy clusters that contain extended radio sources and on a control sample of radio-quiet galaxy clusters. In this way, we will be able to use a larger sample for determining if extended radio sources are more likely to be found in clusters that are undergoing mergers.

\acknowledgments

This work was partially supported by a NASA Long Term Space Astrophysics 
grant NAGW-3152, NASA ROSAT grant NAG5-1819, and NSF 
grant AST93-17596 to JOB. This research has made use of the NASA/IPAC Extragalactic 
Database (NED) which is operated by the Jet Propulsion Laboratory, California Institute 
of Technology, under contract with the National Aeronautics and Space Administration. 
We would like to thank the Steward Observatory
TAC committee for granting time to perform our observations, S. Snowden for the ROSAT 
image processing software, Z. Huang \& C. Sarazin for the AKS software, K. Roettiger and C. Loken for 
fruitful discussions, NOAO for the IRAF package, NRAO for 
AIPS, CfA for PROS, and GSFC for XSPEC. Finally, we thank the referee, W. Oegerle, for a careful reading of this manuscript and for his suggestions that improved the final version of this paper.

\newpage

\newpage
\begin{center}
%References
 {\bf Figure Captions}
\end{center}
\def\count    {counts/s/arcmin$^2$~}

{\bf Fig 1a:}  Overlay of a 20 cm radio map grey scale onto ROSAT X-ray surface brightness contours. The contour map shows the adaptive smoothed X-ray emission from A578. The contour levels are 0.06, 0.10, 0.20, 0.30, 0.50, 0.60, 0.70, 0.80, and 1.00 counts arcmin$^{-2}$.

{\bf Fig 1b:} As in Figure 1a) for the cluster A1569. The contour levels are 0.06, 0.12, 0.24, 0.36, 0.48, 0.60, 0.84, 1.08, and 1.32 counts arcmin$^{-2}$.

{\bf Fig 2a:}  Velocity histogram of all the galaxies observed in the A578 field. The binsize is 200 km/s. We have also marked (crosses) the positions of the radio galaxies.

{\bf Fig 2b:}  2-D velocity-coded plot of all the galaxies observed in the A578 field. The triangles correspond to the galaxies with velocities between 10,000 and 24,000 km/s. The squares correspond to the 24,000-28,000 km/s velocity range. Finally, circles correspond to the galaxies with velocities $>$ 28,000 km/s. The stars mark the positions of the radio galaxies.

{\bf Fig 2c:}  Velocity histogram of the 41 possible galaxy members for A578. We have also marked (crosses) the positions of the radio galaxies.

{\bf Fig 2d:}  2-D velocity-coded plot of the 41 possible galaxy members for A578. The circles represent the galaxies with velocities between 25,000 and 27,000 km/s. The squares show the positions of the galaxies with velocities between 27,000 and 29,000 km/s. Again, the stars mark the position of the radio galaxies. 

{\bf Fig 2e:}  2-D plot of the result of the KMM objective partition of the galaxies.
We have also marked the position of the radio galaxies (stars). The squares correspond to the members of the North group whereas circles represent the galaxies belonging to the South group.

{\bf Fig 2f:}  Overlay of an X-ray map grey scale onto galaxy surface density contours for A578. The continuous contours represent the surface density of the northern subgroup with the following levels: 0.014, 0.028, 0.048, 0.067, 0.084, 0.106, and 0.12 galaxies arcmin$^{-2}$. The dashed contours correspond to the surface density of the southern galaxy subgroup. Their contour levels are 0.01, 0.02, 0.03, 0.04, 0.06, 0.09, 0.1, 0.18, 0.23, 0.29, 0.34, 0.4, and 0.5 galaxies arcmin$^{-2}$. Note how the elongation of the southern subgroup mimic the elongation of the central X-ray emission. 

{\bf Fig 3a:}  Velocity histogram of all the galaxies observed in the A1569 field. The binsize is 200 km/s. We have also marked (crosses) the positions of the radio galaxies.

{\bf Fig 3b:}  Velocity histogram of the 41 possible galaxy members for A1569. We have also marked (crosses) the positions of the radio galaxies.

{\bf Fig 3c:}  2-D velocity-coded plot of all the galaxies observed in the A1569 field. The squares show the positions of the galaxies with velocities between 19,000 and 22,000 km/s. The circles represent the galaxies with velocities between 22,000 and 25,000 km/s. The stars mark the position of the radio galaxies.

{\bf Fig 3d:}  2-D plot of the result of the KMM objective partition of the galaxies.
We have also marked the position of the radio galaxies (stars). The squares correspond to the members of the southern group whereas circles represent the galaxies belonging to the northern group.

{\bf Fig 3e:}  Velocity histogram of the result of the KMM objective partition of the galaxies in A1569. The binsize is 200 km/s. We have also marked the positions of the radio galaxies (squares). The dashed lines represent the velocity distribution of the galaxies that belong to the southern subgroup.

{\bf Fig 3f:}  Overlay of an X-ray map grey scale onto a galaxy surface density contour map. The contour map represents the surface density of all the galaxies within 19,000-25,000 km/s. The contour levels are 0.15, 0.25, 0.4, 0.55, 0.7, and 0.85 galaxies arcmin$^{-2}$. Note how there are two main groups of galaxies that coincide with the two main peaks of X-ray emission. It also appears that these two groups are connected by a filament of X-ray emitting gas and galaxies.

{\bf Fig 4:}  Relative galaxy velocity and number density as a function of $\beta_j$ for the NAT in A578 and the WAT in A1569. The thick lines correspond to the density while the thin lines correspond to the galaxy velocity. The efficiency of the conversion between the jet energy and the radio luminosity is indicated by the percentages.

{\bf Fig 5:}  Overlay of three contour plots. The continuous contour plot represents the gas density from a 4 to 1 mass ratio merger simulation 1 Gyr after core crossing. These contours have the following levels: 1, 2, 3, 4, 5, 6, 7, 8, 9, 10, 11, 12, 13, 14, 15, 16, 17, 18, and 19 times 0.00023 $cm^{-3}$. The long dashed contours represent the surface density of N-body particles that belong to the main cluster with the following contour levels with respect to the peak value: 100\%, 84\%, 42\%, 21\%, and 11\%. This main cluster had a pre-merger mass 4 times greater than the secondary cluster. The short dashed contours represent the surface density of the particles belonging to the secondary cluster with the following contour levels in units of the peak value: 100\%, 77\%, 52\%, and 26\%. Note how the spatial elongations of these two subgroups follow the direction of the merger (from left to right) and how they mimic the distribution of the galaxies in A578. Each tic mark is spaced by 20 kpc.

{\bf Fig 6:}  Histogram of the relative velocity of the gas with respect to the N-body particles for a 4 to 1 merger simulation. The particles correspond to one of the subgroups of the secondary cluster (the rightmost subclump of particles). The time corresponds to 1 Gyr after core crossing. If these particles are similar to the galaxies of the cluster, then this plot shows the effective relative velocity of the gas with respect to the galaxies that can be responsible for the bending of the radio jets (NAT and WAT).

\end{document}